\documentclass[conference]{IEEEtran}
\IEEEoverridecommandlockouts

\usepackage{cite}
\usepackage{graphicx}
\usepackage{textcomp}
\usepackage{xcolor}

\usepackage{graphicx}
\usepackage{comment}
\usepackage{hyperref}
\usepackage[noend]{algpseudocode}
\algrenewcommand\algorithmicindent{0.6em}%
\usepackage{soul}
\setstcolor{red}
\usepackage{threeparttable}
\usepackage{enumitem}
\usepackage{orcidlink}

\usepackage[utf8]{inputenc}
\usepackage{pgfplots}
\DeclareUnicodeCharacter{2212}{−}
\usepgfplotslibrary{groupplots,dateplot}
\usetikzlibrary{patterns,shapes.arrows}

\pgfplotsset{compat=newest}

\usepackage{xspace}
\usepackage{amsmath}
\usepackage{amssymb}
\usepackage{mathtools}
\usepackage{nicefrac}
\usepackage{amsfonts}
\usepackage{pifont}
\usepackage{floatflt}
\usepackage{url}
\usepackage{dsfont}

\usepackage{tikz}
\usetikzlibrary{calc, positioning, shadows, backgrounds}
\usepackage{multirow}
\usepackage{booktabs}
\usepackage{tabularx}
\newcolumntype{L}{>{\raggedright\arraybackslash}X}

\usepackage{algorithm}

\makeatletter
\DeclareRobustCommand*\cal{\@fontswitch\relax\mathcal}
\makeatother

\newcommand{\lime}{\textsc{lime}\xspace}
\newcommand{\plainshap}{{\sc shap}\xspace}

\newcommand{\acfpm}{\textsc{\sc ac}4{\sc pm}\xspace}

\newcommand{\ks}{\textsc{KernelShap}\xspace}

\newcommand{\commentout}[1]{}

\newcommand{\U}{{\cal U}}

\newcommand{\cS}{{\cal S}}

\newcommand{\cF}{{\cal F}}
\newcommand{\V}{{\cal V}}
\newcommand{\R}{{\cal R}}

\newcommand{\sat}{\models}

\newcommand{\lem}{\begin{lemma}}
\newcommand{\elem}{\end{lemma}}
\newcommand{\pro}{\begin{proposition}}
\newcommand{\epro}{\end{proposition}}

\newcommand{\dfn}{\begin{definition}}
\newcommand{\edfn}{\end{definition}}

\newcommand{\xam}{\begin{example}}
\newcommand{\exam}{\end{example}}
\newcommand{\prf}{\noindent{\bf Proof.} }

\usepackage{algorithm}

\usepackage{longtable}
\usepackage{subcaption}
\usepackage{pgfplots}
\usepackage{pgfplotstable}
\usepackage{paralist}
\usepgfplotslibrary{fillbetween}
\pgfplotsset{compat=1.16}

\usepackage[capitalize,noabbrev]{cleveref}

\crefname{line}{line}{lines}
\crefname{figure}{Fig.}{Figs.}
\Crefname{figure}{Fig.}{Figs.}
\crefname{equation}{Eq.}{Eqs.}
\Crefname{equation}{Eq.}{Eqs.}
\crefname{section}{Sec.}{Secs.}
\Crefname{section}{Sec.}{Secs.}
\crefname{definition}{Def.}{Defs.}
\Crefname{definition}{Def.}{Defs.}
\crefname{algorithm}{Alg.}{Algs.}
\Crefname{algorithm}{Alg.}{Algs.}
\Crefname{algocf}{Alg.}{Algs.}
\Crefname{appendix}{Appendix}{Appendices}

\newtheorem{example}{Example}
\newtheorem{lemma}{Lemma}

\newtheorem{definition}{Definition}

\newtheorem{proposition}{Proposition}

\newcommand{\aref}[1]{\hyperref[#1]{Appendix~\ref*{#1}}}

\def\BibTeX{{\rm B\kern-.05em{\sc i\kern-.025em b}\kern-.08em
    T\kern-.1667em\lower.7ex\hbox{E}\kern-.125emX}}

\begin{document}

\title{Causal Explanations of Process Monitor Predictions
}

\author{\IEEEauthorblockN{Tom Yaacov$^{*}$\orcidlink{0000-0002-0565-6506}}
\IEEEauthorblockA{\textit{Dept. of Informatics} \\
\textit{King's College London}\\
London, UK \\
tom.yaacov@kcl.ac.uk}
\and
\IEEEauthorblockN{Nathan Blake\orcidlink{0000-0002-6404-514X}}
\IEEEauthorblockA{\textit{Dept. of Informatics} \\
\textit{King's College London}\\
London, UK \\
nathan.blake@kcl.ac.uk}
\and
\IEEEauthorblockN{Hana Chockler\orcidlink{0000-0003-1219-0713}}
\IEEEauthorblockA{\textit{Dept. of Informatics} \\
\textit{King's College London}\\
London, UK \\
hana.chockler@kcl.ac.uk}
\thanks{* Corresponding Author.}
\thanks{** This work was supported by the Causality in Healthcare AI (CHAI) Hub [UKRI AI and EPSRC grant EP/Y028856/1].}
}

\maketitle

\begin{abstract}
Process mining is widely used to diagnose processes and identify performance and compliance issues. Specifically, Predictive Process Monitoring (PPM) techniques use AI models to predict outcomes of ongoing process instances. While these models can achieve high predictive performance, their black-box nature makes it difficult to understand the underlying reasons behind their output predictions. In this paper, we propose a novel approach for generating local (case-level) explanations of process monitor predictions based on the framework of actual causality. We define a causal model tailored to processes that captures temporal dependencies between events in a trace, thus allowing us to reason about causal influence of events on the predicted outcome. Our method uses this model implicitly to compute causes and quantify the importance of different events with respect to the predicted outcome. We present a practical, model-agnostic algorithm that approximates event responsibility given the process structure reflected in the causal model. We evaluate our approach on a range of datasets derived from real-life event logs from a standard PPM benchmark. Each dataset contains up to 130,000 traces, with trace lengths of up to 1,800 events and up to 400 distinct event types. We compare our approach with state-of-the-art local explanation methods. The results demonstrate that our approach produces more stable and concise explanations while maintaining competitive efficiency.
\end{abstract}


\section{Introduction}
\label{sec:intro}

Process mining techniques analyze event logs from an information system in order to discover, monitor, and ultimately improve the process. 
Predictive Process Monitoring (PPM) is a subfield of process mining that applies machine learning methods to predict future events in the 
process. Deep learning (DL) has substantially improved this predictive capability \cite{ceravolo_predictive_2024}. However, DL models are 
typically opaque, giving no insights into how a prediction was made. Explainable AI (XAI) refers to a family of techniques designed to 
confer some understanding of the workings of DL models, facilitating trust in their predictions \cite{rizzi_explainable_2024}. Post-hoc XAI is one type of XAI which is applied to a model after training and explains a particular instance. These are by far the most commonly 
applied to PPM, with Shapley Additive exPlanations (\plainshap) \cite{shap} and Local Interpretable Model-Agnostic Explanations (\lime) 
\cite{lime} being the most popular methods \cite{kim_illuminating_2025}. However, these are general XAI methods, developed for imaging and 
tabular data, and do not account for the temporal and causal structures inherent to event logs. They are particularly ill-suited to tasks 
involving highly correlated datasets \cite{elkhawaga_explainability_2022}, which is common in PPM domains such as healthcare where understanding causal relationships is crucial. 
The fidelity of XAI techniques is important in any business process, whether this be for profit maximization, or a medico-ethical imperative in fields such as medicine. These necessitate more refined XAI approaches which capture causal structures.


In this proof-of-concept work, we seek to remedy these limitations by using the framework of \emph{actual causality}, introduced by Halpern and Pearl~\cite{HP05a} and extended by Halpern in~\cite{Hal19}. We introduce a causal model for processes that captures temporal dependencies between events in a trace and enables reasoning about their causal influence on process monitor predictions. Using this model, we derive explanations in terms of causes and quantify the importance of different events using the established notion of causal \emph{degree of responsibility}. We further present a model-agnostic algorithm that approximates the causal responsibility of events in the trace for predictions. This algorithm is implemented in the \acfpm tool (Actual Causality for Process Monitoring), which we compare with \lime and \plainshap.

The framework of actual causality is different to \emph{type causality} as introduced by Pearl \cite{pearl_causality_2009}, which deduces causal relationships between the variables based on the existing data, and is forward-looking, that is, used for prediction. In contrast, the study of actual causality is backward-looking (``what caused a certain event to happen?''), in alignment with our interest in this paper. The reader is referred to \Cref{sec:ac} for an overview of actual causality. 

Counterfactual methods have been introduced to address concerns similar to those we consider in this paper \cite{mehdiyev_interpretable_2025}. These XAI methods aim to identify minimal changes to a process instance that would alter the prediction outcome, thereby providing actionable insights for process improvement. This gives practitioners access to 'what-if' examples which are intuitive to grasp. However, certain causal pathways can be particularly nuanced and existing counterfactual methods do not capture all of these. 
We consider four causal behaviors which counterfactuals do not fully capture: \emph{over-determination}, \emph{preemption}, \emph{non-minimal counterfactuals} and \emph{causally inconsistent counterfactuals}.

\begin{figure}[h]
\centering
\begin{tikzpicture}[outer sep=auto, scale=0.85, transform shape]

    \node (v1) at (-2.8, 0) [draw, circle, minimum size=0.9cm, fill=red!10] {$t$};
    \node (v2) at (-1.3, 0.8) [draw, circle, minimum size=0.9cm, fill=red!10] {$a$};
    \node (v3) at (-0.7, -0.8) [draw, circle, minimum size=0.9cm, fill=red!10] {$k$};
    \node (v4) at (2.4, 0) [draw, circle, minimum size=0.9cm, fill=red!10] {$d$};

    \node at (-2.8, -0.8) {\small Blood};
    \node at (-2.8, -1.1) {\small  transfusion};
    \node at (-1.3, 1.5) {\small Anaphylaxis};
    \node at (-0.7, -1.6) {\small Hyperkalaemia};
    \node at (2.4, 0.8) {\small Death};

    \draw[->] (v1) -- (v2);
    \draw[->] (v1) -- (v3);

    \draw[->, very thick] (v2) -- node[above, yshift=2pt] {\small caused} (v4);
    \draw[->, dotted] (v3) -- node[below, align=center] {\small would\\have caused} (v4);

\end{tikzpicture}
\caption{Illustration of preemption. Blood transfusion ($t$) triggers an anaphylactic reaction ($a$) that causes death ($d$). Hyperkalaemia (high serum potassium) would also have caused death, but this alternative pathway was \emph{preempted} by death from anaphylaxis.}
\label{fig:preemption_medical}
\end{figure}
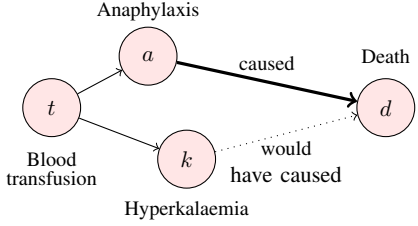

Preemption occurs when a cause first produces an outcome, preventing another potential cause from achieving the same effect. As an example from medicine, consider a patient who receives a blood transfusion. They would die from hyperkalaemia but this is preempted in this case by first dying from an anaphylactic reaction (see \cref{fig:preemption_medical}). A counterfactual explanation might fail to capture that it is the anaphylactic reaction that caused the death, since if the patient had not had the reaction, they would still have died.

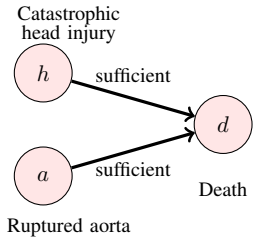
\begin{figure}[h]
\centering
\begin{tikzpicture}[outer sep=auto, scale=0.85, transform shape]

    \node (v2) at (-1.0, 0.8) [draw, circle, minimum size=0.9cm, fill=red!10] {$h$};
    \node (v3) at (-1.0, -0.8) [draw, circle, minimum size=0.9cm, fill=red!10] {$a$};
    \node (v4) at (1.8, 0) [draw, circle, minimum size=0.9cm, fill=red!10] {$d$};

    \node at (-0.6, 1.7) {\small Catastrophic};
    \node at (-0.6, 1.4) {\small head injury};
    \node at (-0.6, -1.6) {\small Ruptured aorta};
    \node at (1.8, -1.0) {\small Death};

    \draw[->, very thick] (v2) -- node[above, yshift=4pt] {\small sufficient} (v4);
    \draw[->, very thick] (v3) -- node[below, yshift=-2pt] {\small sufficient} (v4);

\end{tikzpicture}
\caption{An example of over-determination. Road traffic accident causes both a catastrophic head injury ($h$) and a ruptured aorta ($a$). Either injury alone is sufficient to cause death ($d$), so the outcome is \emph{over-determined}.}
\label{fig:overdetermination_medical}
\end{figure}

Over-determination occurs when multiple \emph{independent} causes exist, any of which on their own are sufficient for an outcome. Consider a patient in a road traffic accident: their catastrophic head injury would be sufficient to lead to death, but so too would a ruptured aorta (see \cref{fig:overdetermination_medical}). A counterfactual explanation might suggest that the two together led to death, without adequately capturing that each alone is sufficient to cause death. This differs from preemption in the temporal ordering of events: in preemption a preceding event brings about an outcome, preempting another event from occurring. In over-determination both sufficient independent events occur simultaneously.

\begin{figure}[h]
\centering
\begin{tikzpicture}[outer sep=auto, scale=0.9, transform shape]

\node (v1) at (-2.4,0) [draw,circle,minimum size=0.9cm,fill=red!10] {$r$};
\node (v2) at (-0.6,0.8) [draw,circle,minimum size=0.9cm,fill=red!10] {$a$};
\node (v3) at (1.8,0) [draw,circle,minimum size=0.9cm,fill=red!10] {$d$};

\node at (-2.4,-0.8) {\small RTA};
\node at (-0.6,1.5) {\small Broken arm};
\node at (1.8,-0.8) {\small Death};

\draw[->,very thick] (v1) -- (v2);
\draw[->,very thick] (v1) -- (v3);

\end{tikzpicture}
\caption{An example of a non-minimal counterfactual: a road-traffic accident (RTA) ($r$) causes both broken arm ($a$) and death ($d$). Clearly, a broken arm is insufficient to cause death. However, the counterfactual trace, $\langle r=0,a=0,d=0 \rangle$, would not be able to discriminate between $r$ and $a$ as the actual cause.}
\label{fig:nonminimal_cf_medical}
\end{figure}
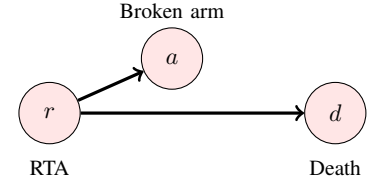

Causally inconsistent counterfactuals occur when a counterfactual is applied which successfully flips a predicted outcome, but violates the underlying causal structure (see \cref{fig:inconsistent_cf_medical}). Improved counterfactual methods generally capture this dependence~\cite{buliga_generating_2025}. Non-minimal counterfactuals occur when changes are made which are not part of the cause. This successfully changes the outcome, but obfuscates the actual causal path that was followed (see \cref{fig:nonminimal_cf_medical}).


    \begin{figure}[h]
        \centering
        \begin{tikzpicture}[outer sep=auto, scale=0.85, transform shape]

            \node (v1) at (-2.8, 0) [draw, circle, minimum size=0.9cm, fill=red!10] {$m$};
            \node (v2) at (-0.6, 0) [draw, circle, minimum size=0.9cm, fill=red!10] {$c$};
            \node (v3) at (1.6, 0) [draw, circle, minimum size=0.9cm, fill=red!10] {$d$};

            \node at (-2.8, -0.9) {\small Multi-organ};
            \node at (-2.8, -1.2) {\small failure};
            \node at (-0.6, -0.9) {\small Cardiac};
            \node at (-0.6, -1.2) {\small arrest};
            \node at (1.6, -0.9) {\small Death};

            \draw[->, very thick] (v1) -- (v2);
            \draw[->, very thick] (v2) -- (v3);

            \draw[->, dashed] (v2) to[bend left=35] node[above] {\small CF: set $c=0$} (v3);

        \end{tikzpicture}
        \caption{An example of a causally-inconsistent counterfactual (CF): a particular patient has multi-organ failure ($m$), which in this case causes cardiac arrest, which causes death ($d$). A counterfactual that sets $c=0$, while leaving $m$ unchanged, may flip the outcome prediction, but it violates the underlying causal structure because the cardiac arrest is itself caused by the organ failure. }
        \label{fig:inconsistent_cf_medical}
    \end{figure}
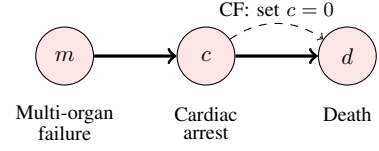

Actual causality extends and generalizes counterfactual (but-for) causality by allowing structured counterfactual reasoning under \emph{contingencies} in causal models. In this setting, an event $A$ is considered an actual cause of $B$ if, possibly under some contingency on other variables, $B$ counterfactually depends on $A$. This framework can accurately capture both preemption and over-determination. 
Moreover, by restricting attention to plausible contingencies within the causal model, it avoids explanations that rely on inconsistent or unrealistic scenarios. Actual causality also allows reasoning about particular instances in the causal model, which helps address issues of non-minimal counterfactual explanations.

We evaluate \acfpm on 22 datasets derived from nine real-life event logs from a widely used predictive process monitoring benchmark. The datasets vary substantially in size and complexity, containing up to 130,000 traces, with trace lengths of up to 1,800 events and up to 400 distinct event types. We compare \acfpm against \lime and \plainshap using established evaluation criteria for explainability, including consistency, succinctness, and computational efficiency. The results show that \acfpm produces explanations that are generally more stable and concise than those of the competing approaches, while remaining computationally practical across all evaluated benchmarks.

The paper structure is as follows. \Cref{sec:preliminaries} introduces the necessary background on process mining and the framework of actual causality. \Cref{sec:causal-model} presents the proposed causal model for processes and discusses the design principles underlying it. \Cref{sec:algorithm} describes our algorithm for approximating the causal responsibility of events in a trace with respect to the prediction of a process monitor. \Cref{sec:evaluation} details an experimental evaluation of the suggested approach on a range of PPM benchmarks, and compares it with state-of-the-art explanation methods. 
We review work related to this paper in \Cref{sec:related-work} and conclude in \Cref{sec:conclusion}.
The full evaluation results and code are available in the supplementary material at \url{https://github.com/ReX-XAI/AC4PM}.

\section{Preliminaries}
\label{sec:preliminaries}

\subsection{Process Mining and Predictive Process Monitoring} 
Process mining comprises a range of techniques for analyzing event data to understand and improve operational processes. In what follows, we briefly present the relevant definition from process mining.

We assume a process instance is a sequence of \emph{activities} taken from a finite set $\mathcal{A}$, allowed in the context of a given business process. An \emph{event} corresponds to the occurrence of an activity as a part of a specific process instance, called a \emph{case}. Events may include additional attributes, such as a timestamp or resources. In this work, we consider a simplified representation where only activity names are recorded for each event, and the timestamps are only used for event ordering. This means that each case can be represented as a sequence of activities, called a \emph{trace} $\sigma = \langle e_1, e_2, \dots,e_n \rangle \in \mathcal{A}^*$, where each event $e_i \in \mathcal{A}$ marks the activity executed at position $i$ in the trace. An \emph{event log} is a collection of traces representing multiple executions of the same process. More formally, an event log is a multiset (since a log can contain multiple instances of the same trace) of traces.

Predictive Process Monitoring (PPM) is a subfield of process mining that aims to predict future aspects of an ongoing business process execution. Examples of prediction targets include the outcome of a process instance, its completion, or its future activities. In this paper, we focus on categorical prediction targets, and therefore assume that the process monitor is a classifier, denoted by $\mathcal{C}$. PPM typically consists of two phases: an offline phase, in which a collection of labeled traces (or prefixes of traces) is used to train $\mathcal{C}$, and an online phase, in which, given a trace (or a prefix of it), $\mathcal{C}$ predicts its label. A comprehensive overview of predictive process monitoring can be found in~\cite{DiFrancescomarino2022}.

\subsection{Actual Causality} 
\label{sec:ac}
In what follows, we informally introduce the relevant concepts from the theory of actual causality. The reader is referred to~\cite{Hal19} for a more in-depth overview.

We assume the world is described in terms of variables and their values. Variables may have a causal influence on others. It is useful to split these variables into two sets: the {\em exogenous\/} variables, $\U$, whose values are determined by factors outside the causal model, and the {\em endogenous\/} variables, $\V$, whose values are dependent on and ultimately determined by the values of exogenous variables.

Formally, a \emph{causal model} is a pair $M = (\cS, \cF)$, where $\cS = (\U, \V, \R)$ is a \emph{signature}, which lists the set of exogenous variables $\U$ and endogenous variables $\V$. $\R$ is a mapping which associates every variable $Y \in \U \cup \V$ with a non-empty set $\R(Y)$ of possible domain values for $Y$. $\cF$ defines a set of structural equations that determine the values of each endogenous variable using variables in $\U \cup \V$. 

A \emph{context} $\vec{u}$ is a setting for the exogenous variables $\U$. 
We call a pair $(M,\vec{u})$ consisting of a causal model $M$ and a context $\vec{u}$, a \emph{causal setting}.
$\vec{u}$ determines, through the structural equations, the values of all endogenous variables. This can be either directly or indirectly, when an endogenous variable depends on other endogenous variables. Following previous literature, we assume that $M$ is recursive (acyclic).  That means that given a context $\vec{u}$, the values of all other variables are determined. We write $(M,\vec{u}) \sat \varphi$ if a Boolean formula $\varphi$ is true in the causal setting $(M,\vec{u})$.

A causal model allows us to perform \emph{interventions} on $(M,\vec{u})$ by changing the value of some variable(s) (subset of $\V$) $X$ to $x$, which essentially amounts to replacing the equation of $X$ in $\cF$ to $X = x$. We denote it by $[X \leftarrow x]$. With that, we can define actual causes:

\dfn[Actual cause~\cite{Hal15,Hal19}]\label{def:AC}
$\vec{X} = \vec{x}$, a subset of the endogenous variables and their valuation, is 
an \emph{actual cause} of $\varphi$ in the causal setting $(M,\vec{u})$ if the following three conditions hold: 
\begin{description}
\item[{\rm AC1.}]\label{ac1} $(M,\vec{u}) \models (\vec{X} = \vec{x})$ and $(M,\vec{u}) \models \varphi $. 
\item[{\rm AC2.}]\label{ac2} There is an alternative setting $\vec{x}'$ of the variables in $\vec{X}$, a 
(possibly empty)  set $\vec{W}$ of variables in $\V - \vec{X}$, and a setting $\vec{w}$ of the variables in $\vec{W}$ such that $(M,\vec{u}) \models \vec{W} = \vec{w}$ and
$(M,\vec{u}) [\vec{X} \gets \vec{x}', \vec{W} \gets
    \vec{w}] \models \neg \varphi$.
\item[{\rm AC3.}] \label{ac3}\index{AC3}  
  $\vec{X}$ is minimal, i.e., there is no strict subset $\vec{X}'$ of
  $\vec{X}$ such that $\vec{X}' = \vec{x}''$ can replace $\vec{X} =
  \vec{x}'$ in 
  AC2, where $\vec{x}''$ is the restriction of
$\vec{x}'$ to the variables in $\vec{X}'$.
\end{description}
\edfn

Essentially, AC1 states that the cause, $\vec{X} = \vec{x}$, and the outcome, $\varphi$ indeed happened in context $\vec{u}$ (this is \emph{actual} causality).
AC2 says that for $\vec{X} = \vec{x}$ to be a cause of $\varphi$, there must be an assignment $\vec{x}'$ for $\vec{X}$ such that if we clamp  $\vec{X} = \vec{x}'$ and  $\vec{W} =\vec{w}$,  $\varphi$ no longer holds. AC3 states that there should be no unnecessary variables in $\vec{X} = \vec{x}$. 
A variable $X$ in an actual cause $\vec{X}$ is called a \emph{part of a cause}. Following~\cite{Hal15}, we often refer to parts of causes as causes.

In this paper, we use the definition of \emph{degree of responsibility}, first introduced in~\cite{CH04}, which quantifies the measure of causal influence.

\dfn[Responsibility \cite{CH04}]\label{def:resp}
The \emph{degree of responsibility} of an endogenous variable $X$ for $\varphi$ in $(M,\vec{u})$ is defined as
$1/(|\vec{X}|+|\vec{W}|)$, where $\vec{X}$ is an actual cause for $\varphi$ containing $X$ that minimizes $|\vec{X}|+|\vec{W}|$, and $\vec{W}$ is its respective contingency.
If there is no actual cause containing $X$, its degree of responsibility is $0$.
\edfn

In the context of explaining the predictions of black-box models, the definition of responsibility provides an ordering of features based on their importance to the predicted value. 


\section{Causal Model for Processes}
\label{sec:causal-model}

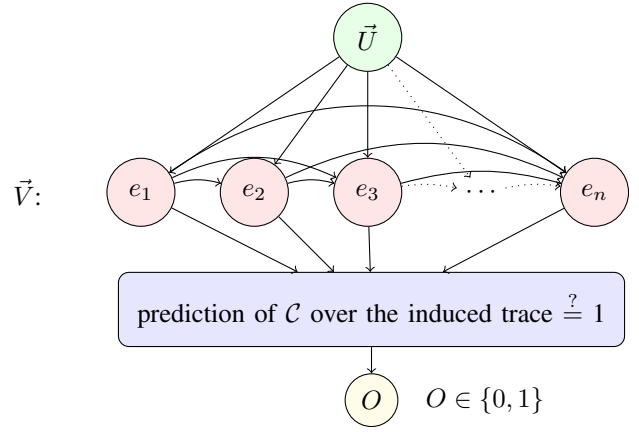
\begin{figure}
    \centering
    \begin{tikzpicture}[outer sep=auto, transform shape]
        \node (u) at (0.75,1.8) [draw, circle, minimum size=0.9cm,fill=green!10] {$\vec{U}$};
        \node (V) at (-3.75, -0.25) {$\vec{V}$:};
        \node (v1) at (-2.25, -0.25) [draw, circle, minimum size=0.9cm,fill=red!10] {$e_1$};
        \node (v2) at (-0.75, -0.25) [draw, circle, minimum size=0.9cm,fill=red!10] {$e_2$};
        \node (v3) at (0.75, -0.25) [draw, circle, minimum size=0.9cm,fill=red!10] {$e_3$};
        \node (dots) at (2.25, -0.25) {$\mathbf{\cdots}$};
        \node (vn) at (3.75, -0.25) [draw, circle, minimum size=0.9cm,fill=red!10] {$e_n$};

        \node (f) at (0.8, -1.8)
            [draw, rectangle, rounded corners, fill=blue!10, inner sep=0.25cm]
            {prediction of $\mathcal{C}$ over the induced trace $\overset{?}{=}$ 1};
        \draw [->] (u) -- (v1);
        \draw [->] (u) -- (v2);
        \draw [->] (u) -- (v3);
        \draw [->] (u) -- (vn);
        
        \draw [->] (v1) -- (f);
        \draw [->] (v2) -- (f);
        \draw [->] (v3) -- (f);
        \draw [->] (vn) -- (f);

        \draw[->] (v1) to[bend left=15] (v2);
        \draw[->] (v1) to[bend left=25] (v3);
        \draw[->] (v1) to[bend left=35] (vn);
        
        \draw[->] (v2) to[bend left=15] (v3);
        \draw[->] (v2) to[bend left=25] (vn);
        
        \draw[->] (v3) to[bend left=15] (vn);
        
        \draw[->, dotted] (v3) to[bend left=10] (dots);
        \draw[->, dotted] (dots) to[bend left=10] (vn);
        \draw[->, dotted] (u) to (dots);

        \node (o) at (0.8, -3.0)
            [draw, circle, minimum size=0.5cm, fill=yellow!10] {$O$};
        \node (oin) at (2.3, -3.0) {$O \in \{0,1\}$};

        \draw [->] (f) -- (o);
    \end{tikzpicture}
    \caption{Graphical representation of our causal model of processes.}
    \label{fig:causal_model}
\end{figure}

We now present our general causal model for processes. Given a process $P$ and its predictive process monitor, $\mathcal{C}$, we define a causal model $M_{P,\mathcal{C}}$ as follows (see \cref{fig:causal_model}). The set $\V = \vec{V} \cup \{O\}$ of endogenous variables consists of a set $\vec{V}$ corresponding to the trace events, $e_1, \dots, e_n$, where $e_i$ is the event in step $i$. $O$ is the output variable, indicating the output of $\mathcal{C}$ over the trace induced by the assignment of $\vec{V}$. The set $\vec{U}$ of exogenous variables captures external factors and sources of randomness that influence the execution of the process. Consequently, once the exogenous variables are assigned with a context $\vec{U}=\vec{u}$, the resulting trace, and therefore the values of $\vec{V}$, can be fully determined.

The general structure selected for the causal model maintains several key principles in the context of causal analysis of processes:
\begin{description}[leftmargin=0.3cm,labelsep=0.1cm,itemsep=2pt,topsep=3pt]
    \item[Temporal aspect:] Previous work that applied the framework of actual causality to explain black-box AI systems considered a depth-2 causal model~\cite{CH24,CKKS26}, similar in structure to that of \cref{fig:causal_model}, except that all variables in $\vec{V}$ are independent of each other. While this assumption is reasonable in other cases and might help in finding causes more efficiently~\cite{CKKS26}, this is not suitable for the case of processes where intervening on one event may affect subsequent events. To capture this temporal dependency, we model each event as causally dependent on its preceding events.
    \item[Natural ordering:] The value of the event $e_i$ as depicted in \Cref{fig:causal_model} may causally depend on the values of the preceding events $e_1,\dots,e_{i-1}$. For instance, $e_3$ might only depend on $e_1$ but not on $e_2$ for some process. The causal model allows dependencies on all preceding events, as this is the most general setting. The temporal ordering defines the set of causal dependencies permitted by our model: an event \emph{may} depend on preceding events, but need not do so.
    \item[Reachability:] In our setting, we only consider interventions that lead to traces 
    that are consistent with the underlying behavior of the process (see \cref{fig:inconsistent_cf_medical} that illustrates a possible inconsistent intervention). While in the theory of actual causality
    interventions that disrupt the causal dependencies are allowed (in fact, this is exactly what
    interventions are), since our only source of information about the causal model is
    the set of traces, we restrict it to consistent (or \emph{reachable}) traces.
    This aligns with the notion of \emph{normality} in actual causality, 
    in which we only consider possible alternatives that can be induced by another context~\cite{Hal08}.
    More specifically, process $P$ defines a set of possible contexts (values of exogenous variables) 
    for the causal model. Each context leads to an actual trace that $P$ can produce. 
    As for the process monitor, we evaluate interventions that remain 
    within the distribution on which it was trained. 
    \item[Simplicity:] The size of the model depends on the length of the trace. However, as we show in \cref{sec:algorithm}, its simple structure allows us to approximate causes and responsibilities without explicitly constructing the full causal model. Furthermore, since each event variable depends on all previous events, we can represent the process using only the event variables themselves, without introducing additional auxiliary variables to keep track of the state during execution.
\end{description}

\section{Computing Responsibilities for Process Outcomes}
\label{sec:algorithm}

In this section, we present and analyze an algorithm that approximates the causal responsibilities of events in the process for the monitor prediction. This is based on the definitions of actual cause and degree of responsibility, as well as the causal model introduced in \Cref{sec:causal-model}. Essentially, the algorithm approximates an importance ranking over the trace events. This ranking can then be used in order to compute explanations, as we show in \autoref{sec:evaluation}.

\subsection{Description of the algorithm} 
The responsibility approximation algorithm is presented in \Cref{alg:responsibility}. It takes as input the trace which we aim to explain, $\sigma = \langle e_1, e_2, \dots,e_n \rangle$, and the predictive process monitor $\mathcal{C}$, whose prediction we aim to explain. 
As we do not have access to the exact causal dependencies and structural equations, the algorithm approximates interventions by sampling traces similar to $\sigma$.
To restrict the set of considered interventions to those that are reachable in the process, we assume access to a process simulator $\mathcal{S}$ capable of generating traces from the process. This simulator may either be provided by the user or discovered from an initial dataset of traces, $\mathcal{D}$, as shown in \cref{sec:evaluation}. The algorithm also receives $\mathcal{D}$ and a budget $k$ specifying the number of samples to generate.

\begin{algorithm}[t]
\caption{Explanation search algorithm}
\label{alg:responsibility}
\renewcommand{\algorithmicrequire}{\textbf{Input:}}
\begin{algorithmic}[1]
        \Require{A trace $\sigma = \langle e_1, e_2, \dots,e_n \rangle$, process monitor $\mathcal{C}$, process simulator $\mathcal{S}$, dataset of traces $\mathcal{D} = \{\sigma_1, \sigma_2, \dots,\sigma_n\}$, number of samples $k$}
    \State $\textit{origPred} \gets \text{predict\_label($\mathcal{C}$,$\sigma$)}$
    \State{\Comment{Sampling using the simulator}}
    \State{\label{line:sample}$\mathcal{D} \gets \mathcal{D} \cup  \text{sample\_k\_traces($\mathcal{S}$,$\sigma$,$k$)}$}
    \State{$p(\sigma_i) \gets \text{predict\_label($\mathcal{C}$,$\sigma_i$)}$ for $\sigma_i$ in $\mathcal{D}$}
    \State{$\textit{resp}(e_i) \gets 0$ for every $e_i$ in $\sigma$}
    \For{\label{line:main-loop}$\sigma_i$ in $\mathcal{D}$}
        \State{\Comment{Check if sample prediction differs from the original}}
        \If{$p(\sigma_i) \ne \textit{origPred}$}
            \State{$S \gets \Call{get\_candidate\_cause}{\sigma, \sigma_i}$}
            \State{$\textit{isMinimal} \gets true$}
            \For{\label{line:inner-loop}$\sigma_i'$ in $\mathcal{D}$}\Comment{Checking minimality }
                \State{$S' \gets \Call{get\_candidate\_cause}{\sigma, \sigma_i'}$}
                \If{$S' \subset S$ and $p(\sigma_i') \ne origPred$}
                    \State{$\textit{isMinimal} \gets false$}
                    \State{\textbf{break}}
                \EndIf
            \EndFor
            \If{$\textit{isMinimal}$}
                \For{$e_i$ in $S$}
                    \State{\Comment{Approximating responsibility}}
                    \State{\label{line:responsibility}$\textit{resp}(e_i) \gets \text{max}(\textit{resp}(e_i), 1/(1+|S|))$}
                \EndFor
            \EndIf
        \EndIf
    \EndFor
    \State{\Return{$\textit{resp}$}}
    \Procedure{\label{line:get-causes}get\_candidate\_cause}{original trace $\sigma$, sampled trace $\sigma'$, dataset of traces $\mathcal{D}$ and their prediction $p$}
    \For{$i$ in $\{1,\dots, |\sigma'|\}$}
        \State{\Comment{Check if prediction is guaranteed from this prefix }}
        \State{\label{line:diff1}$\textit{diff} \gets \{\, \sigma'' \in \mathcal{D} \, \mid \, \sigma'[:i] =\sigma''[:i] \land p(\sigma')\ne p(\sigma'')\,\}$}
        \If{\label{line:diff2}$\textit{diff} = \emptyset$}
            \State{$S \gets \text{get\_set\_of\_changed\_events($\sigma[:i]$,$\sigma'[:i]$)}$}
            \State{\Return{$S$}}
        \EndIf
    \EndFor
    
    \EndProcedure
\end{algorithmic}
\end{algorithm}

The algorithm starts by creating $k$ trace samples using the simulator $\mathcal{S}$ (line~\ref{line:sample}). The sampling process also uses the original trace $\sigma$ to produce samples similar to $\sigma$. If computationally feasible, this step can instead enumerate all possible samples. These samples are added to $\mathcal{D}$, and the process monitor $\mathcal{C}$ is used to compute predictions for all resulting traces. Next, in line~\ref{line:main-loop}, the algorithm iterates over the samples and checks whether any of them changes the prediction of $\mathcal{C}$, corresponding to condition AC2 in \cref{def:AC}. 
When such a sample is found, the algorithm approximates the minimal underlying cause from it using the ``get candidate cause'' procedure (defined from line~\ref{line:get-causes}). This procedure looks for the smallest prefix of the sampled trace from which the change in the prediction is guaranteed in $\mathcal{D}$ (lines \ref{line:diff1}--\ref{line:diff2}). 
In line~\ref{line:diff1}, $\sigma[:i]$ denotes the prefix of $\sigma$ containing its first $i$ events.
The subset of changed events from this prediction-preserving prefix is the approximated candidate cause $S$. Starting from line~\ref{line:inner-loop}, the algorithm then checks the minimality of $S$, corresponding to AC3 in \cref{def:AC}. Specifically, it searches the dataset for evidence that a strict subset of $S$ is enough to change the prediction. If no such subset is found, $S$ is considered a minimal cause. The algorithm then iterates over the events in $S$ and updates their responsibility values whenever the newly computed responsibility exceeds the previously recorded value (line~\ref{line:responsibility}).

\paragraph{\textbf{Computational Considerations}} Let $n$ be the total number of traces in $\mathcal{D}$, including the $k$ additional sampled traces. The dominating part in terms of complexity is the nested loop starting in Line~\ref{line:main-loop}, which calls the ``get candidate cause'' procedure $O(n^2)$ times. When considering the complexity of the ``get candidate cause'' procedure, we get an overall complexity of $O(n^3m^2)$, where $m$ is the maximal trace length. The algorithm makes $O(n)$ calls for prediction from $\mathcal{C}$. 

\subsection{Correctness analysis} 
 
We first prove that if given access to the causal model, the output of \Cref{alg:responsibility} 
satisfies \Cref{def:AC}.
\lem\label{lem:exact}
If \Cref{alg:responsibility} has access to the causal model, its computed cause $S$ is an actual cause of the outcome
according to \Cref{def:AC}. 
\elem
\prf{AC1 is straightforward since it requires the causes and outcome to hold in the original setting, and indeed, we only consider the events that occurred in $\sigma$ as causes of the actual predicted label. Regarding AC2, if given access to the causal model, we can identify the exact set of interventions we consider instead of approximating it using sampling. It follows that we can exactly determine the set of events such that changing them changes the outcome, hence satisfying AC2. The minimality follows a similar line of reasoning: the minimization procedure in the algorithm tries to reduce the candidate set $S$ by checking the effect of removing events from $S$. If given access to the causal model, the algorithm can compute the result directly, and hence accurately minimize the subset cause $S$ given an intervention. Since we have an accurate set of all possible interventions, we can exactly determine if such $S$ is a minimal cause, and therefore satisfy AC3.}
As the causal model is not known, the algorithm instead uses a simulator to produce sample traces. The
traces produced by the simulator are not necessarily for the context of the original input trace. However,
if given access to all traces, we can provide correctness guarantees as shown in the following lemma.
\lem\label{lem:sound}
If \Cref{alg:responsibility} has access to all possible traces of process $P$, its computed cause $S$ satisfies AC1 and AC2 in \cref{def:AC}. 
\elem
\prf{AC1 follows by the same reasoning as in the previous proof, since we consider only events that actually occurred in $\sigma$ as candidate causes of the actual predicted label. Regarding AC2, let $\sigma'$ be the sampled trace from which $S$ is taken, and let $\sigma'[:i]$ be the prefix at which the ``get candidate cause'' procedure ended its execution. $\sigma'[:i]$ contains all elements in $S$. Since the algorithm has access to all possible samples of $P$, we know that the actual continuation of $\sigma'[:i]$ under context $\vec{u}$ in the causal model is one of them. We denote this trace as $\sigma''$. Clearly, the prediction of $\sigma''$ is different from that of $\sigma$, since the algorithm checked that as a part of the ``get candidate cause'' procedure. Therefore, there exists an intervention that changes $\sigma$ to $\sigma''$ and changes the prediction of the outcome. Hence, the algorithm satisfies AC2.
}
Intuitively, AC2 holds because the simulator produces a \emph{superset} of the intervention traces, as it produces the results of all interventions in all possible contexts. It includes, in particular, the results of intervention on the input context, hence the satisfaction of AC2. The minimality condition AC3 may not hold, but the algorithm attempts to approximate minimality by considering subsets of the candidate in its search procedure.

In practice, a complete causal model is not explicitly available, and \Cref{alg:responsibility} relies on a simulator that generates only a subset of all possible traces. Consequently, \Cref{alg:responsibility} can only approximate actual causes and their degrees of responsibility from the sampled traces. The resulting responsibility values should therefore be interpreted as estimates of causal responsibility under the assumed process model and the available samples, rather than as exact degrees of responsibility in a fully specified causal model. The quality of this approximation depends directly on the exhaustiveness of the sample set, which we analyze in \Cref{sec:evaluation}. The experiments show that, in practice, the resulting responsibility ranking is effective in generating stable and concise explanations.


\paragraph*{\textbf{The role of $\vec{W}$}}
In \Cref{def:AC}, the set $\vec{W}$ is a set of variables that are \emph{clamped} to their original values when checking the effect of intervening on $\vec{X}$. Recall that our mechanism for observing the results of interventions is examining the output of a given simulator, which is a probability distribution over the sequence of events. Therefore, it is impossible to measure the precise effect of an intervention or the clamping of a variable. Moreover, clamping the values of variables can result in an unreachable trace. To ensure reachability, the responsibility formula in \Cref{alg:responsibility} assumes that $\vec{W} = \emptyset$. Indeed, as we have only one main causal path (see \Cref{fig:causal_model}), if we intervene on the event $e_i$, then clamping any of the events before it is meaningless (they have their original values anyway), and actively clamping any of the events after $e_i$ could interrupt the propagation of the change or result in an unreachable trace. This is illustrated in \Cref{fig:inconsistent_cf_medical}. This restriction may lead to an overapproximation of some responsibility values. However, as the lemmas show, the resulting algorithm captures actual causes. Furthermore, as demonstrated in \autoref{sec:evaluation}, the resulting responsibility ranking is highly effective in practice for computing explanations.

\section{Experimental Evaluation}
\label{sec:evaluation}

\begin{table*}[ht]
\centering
\caption{Statistics of the datasets used for the experiments.}

\begin{tabular}{l c c c c c c c c c}
\toprule
\bfseries  & \bfseries  & \bfseries  & \bfseries min & \bfseries max & \bfseries mean & \bfseries \#event & \bfseries pos class & \multicolumn{2}{c}{\bfseries test accuracy} \\
\bfseries dataset & \bfseries \#traces & \bfseries \#variants & \bfseries length & \bfseries length & \bfseries length & \bfseries classes & \bfseries  ratio & \bfseries mlp & \bfseries xgboost \\
\midrule
traffic\_fines\_1 & 129615 & 200 & 2 & 20 & 8.12 & 10 & 0.13 & 0.92 & 0.9 \\
sepsis\_cases\_2 & 782 & 707 & 4 & 60 & 14.38 & 15 & 0.11 & 0.97 &  0.93 \\
sepsis\_cases\_4 & 782 & 709 & 4 & 185 & 16.54 & 15 & 0.84 & 0.93 &  0.75 \\
sepsis\_cases\_1 & 782 & 709 & 5 & 185 & 17.36 & 15 & 0.15 & 0.73 &  0.77 \\
hospital\_3 & 77525 & 542 & 2 & 217 & 10.59 & 17 & 0.21 & 0.87 & 0.85 \\
hospital\_2 & 77525 & 1005 & 2 & 217 & 12.39 & 18 & 0.13 & 0.96 &  0.96 \\
Production & 220 & 203 & 1 & 78 & 12.0 & 26 & 0.52 & 0.78 &  0.76 \\
BPIC17\_O\_Cancelled & 31413 & 15846 & 10 & 180 & 48.44 & 26 & 0.41 & 0.99 &  0.96 \\
BPIC17\_O\_Refused & 31413 & 15846 & 10 & 180 & 48.44 & 26 & 0.15 & 1.0 &  0.99 \\
BPIC17\_O\_Accepted & 31413 & 15846 & 10 & 180 & 48.44 & 26 & 0.44 & 0.99 &  0.95 \\
bpic2012\_O\_DECLINED & 4685 & 3790 & 15 & 175 & 43.44 & 36 & 0.17 & 0.99 &  0.97 \\
bpic2012\_O\_ACCEPTED & 4685 & 3790 & 15 & 175 & 43.44 & 36 & 0.54 & 0.99 &  0.93 \\
bpic2012\_O\_CANCELLED & 4685 & 3790 & 15 & 175 & 43.44 & 36 & 0.28 & 0.99 &  0.93 \\
BPIC11\_f3 & 1121 & 811 & 1 & 1368 & 83.15 & 190 & 0.18 & 0.94 &  0.91 \\
BPIC11\_f1 & 1140 & 816 & 1 & 1814 & 77.26 & 193 & 0.36 & 0.87 &  0.85 \\
BPIC11\_f4 & 1140 & 978 & 1 & 1432 & 94.5 & 231 & 0.33 & 0.86 &  0.8 \\
BPIC11\_f2 & 1140 & 978 & 1 & 1814 & 152.44 & 251 & 0.75 & 0.82 &  0.86 \\
BPIC15\_4\_f2 & 577 & 576 & 1 & 82 & 42.07 & 319 & 0.16 & 0.98 &  0.96 \\
BPIC15\_5\_f2 & 1051 & 1049 & 5 & 134 & 52.0 & 376 & 0.31 & 0.99 &  0.98 \\
BPIC15\_1\_f2 & 696 & 677 & 2 & 101 & 42.22 & 380 & 0.23 & 0.93 &  0.94 \\
BPIC15\_3\_f2 & 1328 & 1285 & 3 & 124 & 44.47 & 380 & 0.2 & 0.95 &  0.98 \\
BPIC15\_2\_f2 & 753 & 752 & 1 & 132 & 54.77 & 396 & 0.19 & 0.95 &  0.95 \\
\bottomrule
\end{tabular}
\label{tab:dataset-stats}
\end{table*}

This section evaluates our proposed approach for generating explanations in several case studies. 
To measure the quality of an explanation, we adopt the well-known insertion test~\cite{hama_deletion_2023} which derives a \emph{minimal sufficient} set of features based on their importance (responsibility) ranking, provided by the explanation algorithm. Specifically, features are greedily added based on their order of importance until the selected subset alone is sufficient to reproduce the prediction of the explained instance. That means that sufficiency is satisfied if all instances in the dataset that share the same valuation on this subset receive the same classifier prediction as the explained instance. The size of this sufficient subset serves as a proxy for how accurate the ranking is. Following several principles presented in~\cite{el-khawaga_xai_2022,hama_deletion_2023,CKKS26}, our evaluation concentrates on three main criteria for explanations: 1) Consistency: the explanation algorithm should output similar explanations when executed several times, 2) Efficiency: the explanation algorithm should be efficient and produce an explanation in a reasonable time, and 3) Succinctness: the explanation produced by the algorithm should be as concise as possible. 

\subsection{Case Studies}
We evaluated our approach on a predictive process monitoring benchmark presented in~\cite{teinemaa_outcome-oriented_2019}. This benchmark is based on nine real-life event logs, eight of which are publicly available and one of which is private. The public logs can be accessed through the 4TU Centre for Research Data\footnote{\url{https://www.4tu.nl/htm/research/4tu-research-data}}. From these logs, \cite{teinemaa_outcome-oriented_2019} constructed 22 public case study datasets with binary target labels. Some of the datasets originate from the same underlying event log but differ in their labeling. That is, some datasets represent the same processes but with different outcomes that the monitor is tasked with predicting. 
\cref{tab:dataset-stats} summarizes the main statistics of these datasets, including the number of traces and variants, trace length, number of activity classes, and the label distribution. The table also reports the predictive accuracy of the classifiers used in our experiments.
Each dataset was split into 80\% train and 20\% test.

\subsection{Implementation}
For the evaluation, we implemented \cref{alg:responsibility} in the \acfpm tool. The process simulator used is a Long Short-Term Memory (LSTM) neural network implemented using PyTorch~\cite{paszke_pytorch_2019} and trained to predict the next event given a trace prefix. We pretrained a separate simulator for each case study using its corresponding train dataset. The algorithm input dataset, $\mathcal{D}$, was the train set, and we used $k=100$ additional samples. The complete code for the evaluation and the tool are available at \url{https://github.com/ReX-XAI/AC4PM}.

\subsection{Comparison} 
To assess our contribution, we compared our approach with the current state-of-the-art of local explainability tools for PPM: \lime ~\cite{lime} and \plainshap ~\cite{shap}. Both methods assign importance scores to features, assessing their contribution to the prediction, which aligns with our proposed causal degree of responsibility ranking. \plainshap offers both model-dependent and model-agnostic variants. Since our setting (as well as \lime) assumes a black-box classifier, and we evaluate two different types of classifiers, we selected \ks to ensure a fair and consistent comparison.

\subsection{Experimental Setup} 
The training set was used to train two predictive model classifiers, for which we derived explanations: eXtreme Gradient Boosting (XGBoost)~\cite{chen_xgboost_2016} and a Multi-layer Perceptron (MLP) neural network implemented using PyTorch~\cite{paszke_pytorch_2019}. We used a standard static one-hot encoder for the dataset encoding. The performance of the trained models over the test set is also reported in \cref{tab:dataset-stats}. For each classifier, we compared the results of \acfpm with those of \lime and \plainshap. For the evaluation, we sampled 10 traces from the test set, on which we ran the explanation algorithms. To reduce the effect of poor predictions, we only took samples that the model predicted correctly, that is, true negatives or true positives. No restrictions were placed on trace length, variant frequency, or any other trace characteristics during sampling. For each case, we ran the explanation algorithms 10 times with different random seeds. Experiments were conducted on an Apple M4 Max machine with 32GB RAM. 

\subsection{Evaluation Metrics} 
Based on the evaluation criteria defined above, we assessed the explanation methods using the following metrics. To measure consistency, we computed the Variable Stability Index (VSI)~\cite{visani_statistical_2022}, which quantifies the similarity between explanations output by the insertion tests across multiple runs. For efficiency, we measured the average time required to generate an explanation for a single trace. To evaluate succinctness, we measured the fraction of the explanation out of the total trace.

\subsection{Results}

The results of the local explanation tools for the MLP and the XGBoost classifier are reported in \cref{tab:mlp} and \cref{tab:xgboost}, respectively. The tables present the VSI metric, the explanation size, presented as the fraction of trace elements that appear in the explanation generated using the insertion test, and the average execution time of the explanation algorithm in seconds. \acfpm and \lime successfully completed their runs across all the evaluated datasets. However, \plainshap ran out of memory (\emph{o.m.}) in some settings, even when configured with parameters expected to significantly reduce memory consumption.

\begin{table*}[ht]
\centering
\caption{MLP classifier explanation results}
\begin{threeparttable}
\begin{tabular}{l || c c c || c c c || c c c}
\toprule
\bfseries  & \multicolumn{3}{c ||}{\bfseries VSI}  & \multicolumn{3}{c ||}{\bfseries Explanation size$^1$}  & \multicolumn{3}{c}{\bfseries Execution time$^2$}  \\
\cmidrule(lr){2-4}\cmidrule(lr){5-7}\cmidrule(lr){8-10}
\bfseries Dataset & \bfseries AC4PM & \bfseries LIME & \bfseries SHAP & \bfseries AC4PM & \bfseries LIME & \bfseries SHAP & \bfseries AC4PM & \bfseries LIME & \bfseries SHAP \\
\midrule
traffic\_fines\_1 & \bfseries 0.99 & 0.89 & 0.93 & \bfseries 0.15 & 0.22 & 0.22 & \bfseries 0.03 & 0.06 & 3.39 \\
sepsis\_cases\_2 & 0.91 & 0.54 & \bfseries 1.0 & 0.22 & 0.13 & \bfseries 0.07 & \bfseries 0.22 & 4.43 & 2.43 \\
sepsis\_cases\_4 & 0.89 & \bfseries 0.96 & 0.78 & \bfseries 0.26 & 0.43 & 0.37 & \bfseries 1.13 & 2.33 & 2.19 \\
sepsis\_cases\_1 & 0.77 & 0.63 & \bfseries 0.9 & 0.23 & 0.4 & \bfseries 0.07 & 0.97 & \bfseries 0.62 & 1.73 \\
hospital\_3 & 0.86 & 0.82 & \bfseries 0.97 & \bfseries 0.2 & 0.47 & 0.39 & \bfseries 1.17 & 3.81 & 2.69 \\
hospital\_2 & 0.8 & \bfseries 0.92 & 0.84 & 0.27 & 0.83 & \bfseries 0.25 & 1.22 & \bfseries 0.85 & 2.63 \\
Production & 0.82 & 0.78 & \bfseries 0.91 & 0.18 & 0.53 & \bfseries 0.17 & \bfseries 0.27 & 0.76 & 0.79 \\
BPIC17\_O\_Cancelled & \bfseries 0.99 & 0.85 & \emph{o.m.} & \bfseries 0.33 & 0.63 & \emph{o.m.} & 83.75 & \bfseries 1.85 & \emph{o.m.} \\
BPIC17\_O\_Refused & \bfseries 0.98 & 0.75 & \emph{o.m.} & \bfseries 0.16 & 0.63 & \emph{o.m.} & 55.17 & \bfseries 2.55 & \emph{o.m.} \\
BPIC17\_O\_Accepted & \bfseries 0.99 & 0.53 & \emph{o.m.} & \bfseries 0.12 & 0.32 & \emph{o.m.} & 83.65 & \bfseries 2.25 & \emph{o.m.} \\
bpic2012\_O\_DECLINED  & \bfseries 0.97 & 0.76 & \emph{o.m.} & \bfseries 0.18 & 0.61 & \emph{o.m.} & 11.57 & \bfseries 2.13 & \emph{o.m.} \\
bpic2012\_O\_ACCEPTED & \bfseries 0.96 & 0.44 & \emph{o.m.} & \bfseries 0.27 & 0.37 & \emph{o.m.} & 26.82 & \bfseries 2.15 & \emph{o.m.} \\
bpic2012\_O\_CANCELLED & \bfseries 0.98 & 0.63 & \emph{o.m.} & \bfseries 0.17 & 0.5 & \emph{o.m.} & 15.64 & \bfseries 2.26 & \emph{o.m.} \\
BPIC11\_f3 & \bfseries 0.9 & 0.36 & \emph{o.m.} & \bfseries 0.04 & 0.31 & \emph{o.m.} & 191.19 & \bfseries 19.52 & \emph{o.m.} \\
BPIC11\_f1 & \bfseries 0.93 & 0.56 & \emph{o.m.} & \bfseries 0.1 & 0.3 & \emph{o.m.} & 321.35 & \bfseries 16.97 & \emph{o.m.} \\
BPIC11\_f4 & \bfseries 0.96 &  0.41 & \emph{o.m.} & \bfseries 0.15 & 0.25 & \emph{o.m.} & 312.27 & \bfseries 21.36 & \emph{o.m.} \\
BPIC11\_f2 & \bfseries 1.0 & 0.28 & \emph{o.m.} & \bfseries 0.02 & 0.24 & \emph{o.m.} & 571.42 & \bfseries 28.71 & \emph{o.m.} \\
BPIC15\_4\_f2 & 0.72 & \bfseries 0.81 & 0.23 & \bfseries 0.06 & 0.83 & 0.11 & \bfseries 1.16 & 3.43 & 22.59 \\
BPIC15\_5\_f2 & \bfseries 0.73 & 0.59 & \emph{o.m.} & \bfseries 0.07 & 0.65 & \emph{o.m.} & 5.87 & \bfseries 5.38 & \emph{o.m.} \\
BPIC15\_1\_f2 & 0.8 & \bfseries 0.83 & 0.34 & \bfseries 0.09 & 0.89 & 0.14 & \bfseries 1.74 & 4.65 & 47.63 \\
BPIC15\_3\_f2 & \bfseries 0.81 & 0.7 & \emph{o.m.} & \bfseries 0.04 & 0.77 & \emph{o.m.} & 6.86 & \bfseries 5.22 & \emph{o.m.} \\
BPIC15\_2\_f2 & 0.7 & \bfseries 0.77 & 0.28 & \bfseries 0.11 & 0.76 & 0.2 & \bfseries 4.42 & 5.79 & 78.35 \\
\bottomrule
\end{tabular}
\begin{scriptsize}
$^1$ fraction out of the total trace, $^2$ in seconds, \emph{o.m.} out of memory
\end{scriptsize}
\end{threeparttable}

\label{tab:mlp}
\end{table*}

\begin{table*}[ht]
\centering
\caption{XGBoost classifier explanation results}
\begin{threeparttable}
\begin{tabular}{l || c c c || c c c || c c c}
\toprule
\bfseries  & \multicolumn{3}{c ||}{\bfseries VSI}  & \multicolumn{3}{c ||}{\bfseries Explanation size$^1$}  & \multicolumn{3}{c}{\bfseries Execution time$^2$}  \\
\cmidrule(lr){2-4}\cmidrule(lr){5-7}\cmidrule(lr){8-10}
\bfseries Dataset & \bfseries AC4PM & \bfseries LIME & \bfseries SHAP & \bfseries AC4PM & \bfseries LIME & \bfseries SHAP & \bfseries AC4PM & \bfseries LIME & \bfseries SHAP \\
\midrule
traffic\_fines\_1 & \bfseries 0.98 & 0.97 & \bfseries 0.98 & \bfseries 0.16 & 0.25 & 0.18 & \bfseries 0.03 & 0.04 & 0.9 \\
sepsis\_cases\_2 & 0.86 & 0.64 & \bfseries 1 & 0.22 & 0.08 & \bfseries 0.07 & \bfseries 0.21 & 0.85 & 3.24 \\
sepsis\_cases\_4 & 0.8 & 0.74 & \bfseries 0.98 & 0.17 & \bfseries 0.13 & \bfseries 0.13 & \bfseries 1.33 & 1.92 & 4.26 \\
sepsis\_cases\_1 & 0.86 & 0.75 & \bfseries 1 & 0.12 & 0.29 & \bfseries 0.09 & \bfseries 1.06 & 1.36 & 5.98 \\
hospital\_3 & 0.82 & 0.89 & \bfseries 0.98 & \bfseries 0.19 & 0.5 & 0.4 & \bfseries 1.16 & 1.76 & 2.91 \\
hospital\_2 & 0.83 & \bfseries 0.95 & 0.92 & 0.27 & 0.78 & \bfseries 0.21 & \bfseries 1.3 & 1.72 & 7.24 \\
Production & 0.83 & \bfseries 0.97 & 0.94 & 0.16 & 0.45 & \bfseries 0.1 & \bfseries 0.28 & 1.35 & 3.79 \\
BPIC17\_O\_Cancelled & \bfseries 0.99 & 0.93 & \emph{o.m.} & \bfseries 0.28 & 0.51 & \emph{o.m.} & 83.01 & \bfseries 3.79 & \emph{o.m.}\\
BPIC17\_O\_Refused & \bfseries 0.97 & 0.88 & \emph{o.m.} & \bfseries 0.16 & 0.53 & \emph{o.m.} & 55.4 & \bfseries 3.95 & \emph{o.m.}\\
BPIC17\_O\_Accepted & \bfseries 0.99 & 0.72 & \emph{o.m.} & \bfseries 0.13 & 0.45 & \emph{o.m.} & 88.04 & \bfseries 7.7 & \emph{o.m.}\\
bpic2012\_O\_DECLINED & 0.92 & \bfseries 0.98 & \emph{o.m.} & \bfseries 0.13 & 0.43 & \emph{o.m.} & 11.77 & \bfseries 3.24 & \emph{o.m.}\\
bpic2012\_O\_ACCEPTED & \bfseries 0.99 & 0.78 & \emph{o.m.} & 0.26 & \bfseries 0.12 & \emph{o.m.} & 25.89 & \bfseries 2.98 & \emph{o.m.}\\
bpic2012\_O\_CANCELLED & \bfseries 0.98 & 0.91 & \emph{o.m.} & \bfseries 0.16 & 0.39 & \emph{o.m.} & 15.39 & \bfseries 3.31 & \emph{o.m.}\\
BPIC11\_f3 & 0.92 & \bfseries 0.93 & \emph{o.m.} & \bfseries 0.04 & 0.18 & \emph{o.m.} &  191.41 & \bfseries 26.3 & \emph{o.m.}\\
BPIC11\_f1 & 0.86 & \bfseries 0.88 & \emph{o.m.} & \bfseries 0.14 & 0.38 & \emph{o.m.} & 348.04 & \bfseries 23.02 & \emph{o.m.} \\
BPIC11\_f4 & \bfseries 0.95 & 0.84 & \emph{o.m.} & \bfseries 0.13 & 0.24 & \emph{o.m.} & 232.75 & \bfseries 29.34 & \emph{o.m.}\\
BPIC11\_f2 & \bfseries 0.97 & 0.85 & \emph{o.m.} & \bfseries 0.03 & 0.1 & \emph{o.m.} & 680.93 & \bfseries 40.55 & \emph{o.m.}\\
BPIC15\_4\_f2 & 0.79 & \bfseries 0.92 & 0.77 & \bfseries 0.06 & 0.17 & 0.1 & \bfseries 1 & 4.99 & 49.61 \\
BPIC15\_5\_f2 & 0.74 & \bfseries 0.85 & \emph{o.m.} & \bfseries 0.07 & 0.21 & \emph{o.m.} & \bfseries 5.9 & 8.73 & \emph{o.m.}\\
BPIC15\_1\_f2 & 0.75 & \bfseries  0.85 &0.8 & 0.12 & 0.34 & \bfseries 0.11 & \bfseries 2.64 & 6.74 & 94.87 \\
BPIC15\_3\_f2 & 0.79 & \bfseries 0.94 & \emph{o.m.} & \bfseries 0.04 & 0.26 & \emph{o.m.} & \bfseries 6.88 & 7.72 & \emph{o.m.}\\
BPIC15\_2\_f2 & 0.79 & 0.94 & \bfseries 0.97 & 0.08 & 0.3 & \bfseries 0.04 & \bfseries 3.94 & 8.81 & 216.64 \\
\bottomrule
\end{tabular}
\begin{scriptsize}
$^1$ fraction out of the total trace, $^2$ in seconds, \emph{o.m.} out of memory
\end{scriptsize}
\end{threeparttable}
\label{tab:xgboost}
\end{table*}

For the VSI metric, \acfpm achieves the highest stability in the majority of datasets across both classifiers. In the case of XGBoost, \lime appears slightly more stable than \acfpm. However, we argue that this effect is mainly due to variability in \lime's performance rather than a decrease in the stability of \acfpm: when computing Spearman's correlation~\cite{schober_correlation_2018} between the VSI scores of each method across the two classifiers, \lime shows a relatively low correlation ($\rho \approx 0.43$), whereas \acfpm exhibits a much higher correlation ($\rho \approx 0.88$). Importantly, the classifiers themselves did exhibit significant differences in predictive performance, as reflected by their test accuracies in \cref{tab:dataset-stats}. This suggests that, although \lime is model-agnostic, its stability is highly affected by the type of classifier used, whereas \acfpm (and \plainshap) are more consistent.

Another interesting aspect is the effect of dataset attributes on the stability of the explanation algorithms. We observed that the stability of \acfpm generally increases with the number of trace variants in the dataset, with moderate positive Spearman correlations between these variables (MLP: $\rho \approx 0.44$, XGBoost: $\rho \approx 0.49$). This is largely expected, as more data reduces the variability introduced by additional intervention samples. A similar trend can be observed for \plainshap. In contrast, the stability of \lime generally decreases as the number of variants grows (MLP $\rho \approx -0.32$, XGBoost $\rho \approx -0.09$). A similar pattern appears when examining the effect of trace length. The stability of \acfpm increases with the mean trace length (MLP $\rho \approx 0.22$, XGBoost $\rho \approx 0.18$), whereas the stability of \lime decreases (MLP $\rho \approx -0.66$, XGBoost $\rho \approx -0.16$). The poor stability of \lime is expected. \lime explains predictions by fitting a local surrogate model to randomly generated perturbations of the input. In PPM, the dimensionality of the one-hot encoded representation grows with both the number of activities and the trace length, exacerbating the curse of dimensionality. As a result, the surrogate model becomes increasingly sensitive to the specific perturbation samples used, leading to lower explanation stability~\cite{visani_statistical_2022}.

The explanation size measures the fraction of trace elements sufficient to generate the prediction, and thus reflects how succinct the explanation is. Across the evaluated datasets, explanations derived using \acfpm are consistently smaller than those produced by \lime and \plainshap. This suggests that \acfpm tends to identify a more focused subset of events that is sufficient to reproduce the prediction. Further, the size of \acfpm explanations appears to \emph{decrease} as process complexity increases, both with respect to the number of activities (MLP $\rho \approx -0.71$, XGBoost $\rho \approx -0.68$) and the mean trace length (MLP $\rho \approx -0.53$, XGBoost $\rho \approx -0.52$). This suggests that \acfpm produces more focused explanations in processes with greater variability. In such settings, predictions are likely driven by more specific patterns, allowing \acfpm to isolate smaller sets of causal events. In contrast, we did not observe a similar pattern for \lime and \plainshap, whose correlations did not indicate a consistent direction.

Across the evaluated datasets, the runtime of all methods increases with the complexity of the underlying process data, particularly with respect to the number of traces, their length, and the overall number of activities. This especially affected the execution of \acfpm, whose runtime grows substantially as traces become longer and the number of traces increases. This behavior is consistent with the complexity analysis presented in \autoref{sec:algorithm}. Despite this dependency, the explanation algorithms generally remain relatively efficient, producing explanations within a few minutes across all evaluated benchmarks. That said, the current, proof-of-concept implementation of the \acfpm approach has considerable room for optimization. We therefore expect that the runtime of the algorithms can be significantly reduced in the future. 

Different aspects of dataset complexity also affected the runtime of the compared explanation methods. \lime appeared particularly sensitive to longer traces, whereas \plainshap was more strongly affected by datasets with a larger number of distinct activities. This behavior is expected, as both trace length and the number of activities directly influence the dimensionality of the one-hot encoding, substantially increasing the computational cost of these methods.

\section{Related Work}
\label{sec:related-work}

Several studies have explored the use of explainable AI techniques in the context of process mining. Some have investigated the applicability of general explanation tools such as \lime and \plainshap for interpreting predictions of PPM models and understanding the influence of process attributes and activities on outcomes~\cite{elkhawaga_explainability_2022,el-khawaga_xai_2022,rizzi_explainable_2024}. These approaches typically provide feature-based explanations but often treat process traces in the same format as they were fed into the PPM model, usually in the form of tabular data, ignoring the temporal and structural dependencies processes exhibit.

Another line of work investigates counterfactual explanations for PPM predictions~\cite{huang_counterfactual_2022,9576881}. 
A more recent work has explored counterfactual explanations for process predictions, highlighting several important features of processes in the context of counterfactual trace generation, such as temporal ordering~\cite{buliga_guiding_2025,buliga_generating_2025}. 
Similarly,~\cite{qafari_case_2021} propose a framework for case-level counterfactual reasoning over event logs using structural equation models. While counterfactual explanations may provide valuable insights in some cases, they do not accurately capture causal relationships in more complex cases, as discussed in the introduction.

More broadly, several approaches have explored the use of causal analysis in process mining~\cite{hompes_discovering_2017,leemans_causal_2022,houdt_aitia-pm_2023,waibel_causal_2023,fournier2025business,qafari_root_2020}. These works primarily focus on global (process-level) causal discovery and analysis, rather than local (case-level) explanations for predictions of black-box PPM models, which is the focus of this paper.

Actual causality is well-suited for providing explanations for events that already occurred~\cite{Hal19}. 
It has been used to explain AI models across various domains, such as images~\cite{CKKS26,kelly2026generating}, audio~\cite{kelly_i_2026}, vibrational spectroscopy~\cite{blake2026causal}, and failures of autonomous cyber-physical systems~\cite{icra}. These methods approximate causal responsibility to efficiently find causal explanations of different sizes, confidences, and multiplicities~\cite{CKK25,kelly_i_2025}.
These papers assume causal independence between the inputs and the AI model as a black box, and work
with depth-2 causal models. In contrast, in this work, we take into account the causal dependencies between
the variables, hence our causal models have a higher depth.

\section{Conclusion}
\label{sec:conclusion}

In this paper, we presented a novel approach for generating local explanations for predictions produced by Predictive Process Monitoring (PPM) models using the framework of actual causality. We introduced a causal model tailored to processes that captures the temporal dependencies between events in the trace and enables reasoning about their influence over the predicted outcome. Using this model, we proposed an algorithm that approximates the causal responsibility of events in a trace in order to estimate their importance to the predicted outcome. We evaluated the algorithm on a range of datasets from a widely used PPM benchmark and compared it with state-of-the-art local explanation approaches. The experimental results demonstrate that our proposed approach produces explanations that are generally more stable and concise, while remaining computationally practical. Moreover, for larger sample sets, the explanations computed by our algorithms are more precise, and the approach is stable with respect to the complexity of the traces.

As this is the first step towards applying the framework of actual causality to explain PPM models, we focused on a simplified representation of process traces, where each event is described by its activity name and timestamps are used solely for event ordering. While this setting suits many use cases, processes often include richer information, such as continuous process attributes or contextual event data. Extending the framework to support richer process representations is therefore an important direction for future work. We believe that such an extension is feasible, as previous work has successfully applied actual causality to explain an AI system operating in a high-dimensional input space~\cite{blake2026causal,kelly_i_2026}. Future research will also focus on algorithmic solutions that account for the size of the contingency set when approximating responsibility. This will enable more accurate estimates of responsibility and more faithful rankings. Future work will also investigate evaluation against ground-truth explanations.



\bibliographystyle{IEEEtran}
\bibliography{more}

\end{document}